\documentclass{appolb}
\usepackage{slashed}
\usepackage{epsfig}

\begin{document}

\title{A note on the anomalous magnetic moment of the muon}

\author{Davor Palle
\address{
Zavod za teorijsku fiziku, Institut Rugjer Bo\v skovi\' c \\
Bijeni\v cka cesta 54, 10000 Zagreb, Croatia}
}

\maketitle

\begin{abstract}
The anomalous magnetic moment of the muon is an important observable
that tests radiative corrections of all three observed local gauge
forces: electromagnetic, weak and strong interactions.
High precision measurements reveal some discrepancy with
the most accurate theoretical evaluations of the anomalous magnetic moment.
We show in this note that the UV finite theory cannot resolve this
discrepancy. We believe that more reliable estimate of the nonperturbative 
hadronic contribution and the new measurements can resolve the problem.
\end{abstract}


\section{Introduction and motivation}

Since the most reliable Standard Model (SM) evaluations of the anomalous magnetic moment
of the muon differs three to four standard deviations from the measurements \cite{Blum},
it might be useful to examine predictions for this observable
of the theories beyond the SM.

It is well known that the SM has three major obstacles: (1) massless neutrinos,
(2) absence of a dark matter particle, and (3) absence of the lepton and baryon number
violations. We show in ref. \cite{Palle1} that the UV nonsingular theory, free of
 the $SU(2)$ global anomaly (called the BY theory in \cite{Palle1}), is free of 
the SM deficiencies.
Besides three heavy
Majorana neutrinos as cold dark matter particle(s), the BY theory contains three
light Majorana neutrinos with a relation on the three mixing angles, 
alowing the inclusion of 
the lepton CP violating phase. The effects of the universal UV spacelike cut-off
are studied in various strong and electroweak processes in refs. \cite{Palle2,Palle3,Palle4}.
The cosmological consequences of the nonsingular Einstein-Cartan theory of gravity can be
envisaged  in ref. \cite{Palle5}. The minimal distance in the Einstein-Cartan 
cosmology is compatible with the UV cut-off of the BY theory.

In this note we want to inspect the impact of the UV cut-off on
the radiative corrections of the anomalous magnetic moment
of the muon. Next chapter deals with the explicit calculations, however the relevant
detailed formulae the reader can find in the Appendix. The concluding section
discusses the numerical results and their consequences.

\section{One loop corrections within the UV nonsingular theory}

The anomalous magnetic moment of the muon is one of the most precisely measured 
observable to the uncertainty of ${\cal O}(10^{-9})$ \cite{Blum}. It is calculated 
within the perturbative quantum field theory to the very high order.

The perturbation theory works well and it is very accurate because of the very small fine structure
constant. The anomalous magnetic moment formula can be cast into transparent perturbation 
series where it is possible to compare and study any modification of the Standard Model (SM)
\cite{Gnendiger}.
We evaluate one loop corrections with virtual electroweak bosons within the UV finite BY theory \cite{Palle1}.
Although this theory contains Majorana light neutrinos, we can safely neglect their Majorana character
and their masses because of their smallness. The heavy Majorana neutrinos are coupled strongly to Nambu-Goldstone
scalars at tree level, but not to electroweak gauge bosons. They are, therefore, decoupled in the evaluation
of the anomalous magnetic moment.  

Since the gauge invariant physical result cannot depend on the choice of gauge, we can freely
perform our calculations in the 't Hooft-Feynman gauge with Nambu-Goldstone scalars instead
in the unitary gauge where Nambu-Goldstone scalars are decoupled from lepton doublets \cite{Fujikawa}.
However, the scalar doublet does not contain the Higgs scalar \cite{Palle1} since the noncontractible
space is a symmetry breaking mechanism in the BY theory and the UV cut-off is fixed at tree level
by the weak boson mass $\Lambda = \frac{\pi}{\sqrt{6}}\frac{2}{g} M_{W}$, 
 $e = g \sin \Theta_{W}$, $\cos \Theta_{W} = \frac{M_{W}}{M_{Z}}$ \cite{Palle1}.
The Higgs scalar is decoupled from other particles in the BRST transformations \cite{Kugo} and
does not play essential role in the proof of the renormalizability of the spontaneously
broken gauge theories \cite{Aoki}. Consequently, the UV finite BY theory without the Higgs scalar
is also renormalizable.

It is necessary to comment the claim that the recently discovered $125\ GeV$ scalar resonance is the SM
Higgs scalar \cite{CERN2012}. 
P. Cea \cite{Cea} proposed the most natural explanation of the $125\ GeV$ resonance as
a mixture of toponium and gluonium.
The possible new 750 GeV heavy boson resonance at the LHC \cite{CERN2015} might be a perfect candidate
for a heavier scalar twin of the $125\ GeV$ boson \cite{twins}.

Let us go back to the description of the electroweak one loop contributions to the anomalous 
magnetic moment.
The SM one loop result takes the form \cite{Blum,Gnendiger,Fujikawa,Jeger}
($m$ denotes the mass of the muon and $s_{w}=\sin \Theta_{W}$):

\begin{eqnarray*}
a_{\mu} \equiv \frac{1}{2}(g_{\mu}-2), 
\end{eqnarray*}
\begin{eqnarray*}
a_{\mu}^{\gamma} = \frac{1}{2}\frac{\alpha}{\pi} + {\cal O}((\frac{\alpha}{\pi})^{2})
\end{eqnarray*}
\begin{eqnarray*}
a_{\mu}^{EW(1)} = \frac{G_{F}}{\sqrt{2}}\frac{m^{2}}{8 \pi^{2}}
\{ \frac{10}{3} + \frac{1}{3}(1 - 4 s_{w}^{2})^{2} -\frac{5}{3} + {\cal O}(10^{-5}) \}.
\end{eqnarray*}

We have to rederive these results
equiped with the SM Feynman rules in the 't Hooft-Feynman gauge. One can 
straightforwardly extract the contributions with one virtual photon \cite{Schwinger} and 
one virtual W and Z bosons \cite{Fujikawa} in terms of the coefficient functions of the tensor and vector Green functions that
can be deduced from the scalar Green functions-master integrals \cite{Logan} 
(for definitions and explicit expressions see the Appendix):

\begin{eqnarray}
\lim_{\Lambda \rightarrow \infty} m^{2}(C_{11}+C_{21})(photon) = \frac{1}{2}, 
\end{eqnarray}
\begin{eqnarray}
\lim_{\Lambda \rightarrow \infty} \frac{1}{4}(C_{11}-C_{21})(W\ boson) = \frac{10}{3}\frac{1}{16 M_{W}^{2}}.
\end{eqnarray}
\begin{eqnarray}
\lim_{\Lambda \rightarrow \infty} &\{& \frac{1}{2}[\frac{1}{4}(1-2 s_{w}^{2})^{2}+s_{w}^{4}]
(2 C_{0}+ 3 C_{11} + C_{21}) + s_{w}^{2}(1-2 s_{w}^{2}) \nonumber \\ & &\times (C_{0} + C_{11}) \}  
(Z\ boson) = \frac{1}{48 M_{Z}^{2}}[(3-4 c_{w}^{2})^{2} - 5], 
\end{eqnarray}

The deviation between the SM and the BY theory could be found in the scalar master integrals. The UV cut-off $\Lambda$
in the spacelike domain of the Minkowski spacetime is introduced as a Lorentz and gauge invariant quantity.
The analytical continuation to the timelike domain is performed on the Riemann's sheets, when necessary.
By the symmetrization of the external momenta of the master integrals with the UV cut-off we insure
their translational invariance. Consequently, all the master integrals of the BY theory
have a correct limit $\Lambda \rightarrow  \infty$ of the standard QFT master integrals.

We left our presentation of the explicit results and comments to the next section.

\section{Results and conclusions}

The one loop results for the BY theory follow from Eqs. (1-3) and the expansion formulas for scalar one, two and three
point functions in the SM and the BY theory presented in the Appendix.
Besides the difference between the SM and BY in the cut-off ($\Lambda = \infty$ or 
 $\Lambda = \frac{\pi}{\sqrt{6}}\frac{2}{g} M_{W}$), 
one has to exclude
the Higgs boson contribution for the BY theory ($M_{H}\rightarrow \infty$)\cite{Higgs}:

\begin{eqnarray}
a_{\mu}^{\gamma}(BY)^{(1)} = (\frac{1}{2}-\frac{1}{4}\frac{m^{2}}{\Lambda^{2}}
+\frac{5}{12}\frac{m^{6}}{\Lambda^{6}})\frac{\alpha}{\pi},     
\end{eqnarray}

\begin{eqnarray}
a_{\mu}^{W}(BY)^{(1)} = \frac{G_{F}}{\sqrt{2}}\frac{m^{2}}{8 \pi^{2}} \{ \frac{10}{3} 
- \frac{41}{6} \frac{M_{W}^{2}}{\Lambda^{2}} + \frac{99}{8} \frac{M_{W}^{4}}{\Lambda^{4}}\},
\end{eqnarray}

\begin{eqnarray}
a_{\mu}^{Z}(BY)^{(1)} =\frac{G_{F}}{\sqrt{2}}\frac{m^{2}}{8 \pi^{2}} \{
\frac{1}{3}(1 - 4 s_{w}^{2})^{2} -\frac{5}{3}
+ (\frac{3}{2}+2 s_{w}^{2}-4 s_{w}^{4})\frac{M_{Z}^{2}}{\Lambda^{2}} \nonumber \\
+\frac{1}{48} (43 -332 s_{w}^{2} + 664 s_{w}^{4})\frac{M_{Z}^{4}}{\Lambda^{4}} \}.
\end{eqnarray}

We evaluate numerically the difference with the following set of the well established
parameters (see also the figure for the cut-off dependence): 

\begin{eqnarray*}
m(muon)=105.658\ MeV,\ M_{Z}=91.188\ GeV,\ G_{F}=1.166\times10^{-5}\ GeV^{-2}, \\
\alpha = 1/137.036,\ M_{W}=80.385\ GeV,\ s_{w}^{2}=0.22295,\ 
\Lambda = 326.2\ GeV,
\end{eqnarray*}
\begin{eqnarray*}
& &[a_{\mu}^{BY}-a_{\mu}^{SM}]^{\gamma,1\ loop}=-6.09\times 10^{-11}, \\
& &[a_{\mu}^{BY}-a_{\mu}^{SM}]^{W,1\ loop}=-4.307\times 10^{-10}, \\
& &[a_{\mu}^{BY}-a_{\mu}^{SM}]^{Z,1\ loop}=+1.595\times 10^{-10}, \\
& &[a_{\mu}^{BY}-a_{\mu}^{SM}]^{\gamma+W+Z,1\ loop}=-3.321\times 10^{-10}, 
\end{eqnarray*}

\epsfig{figure=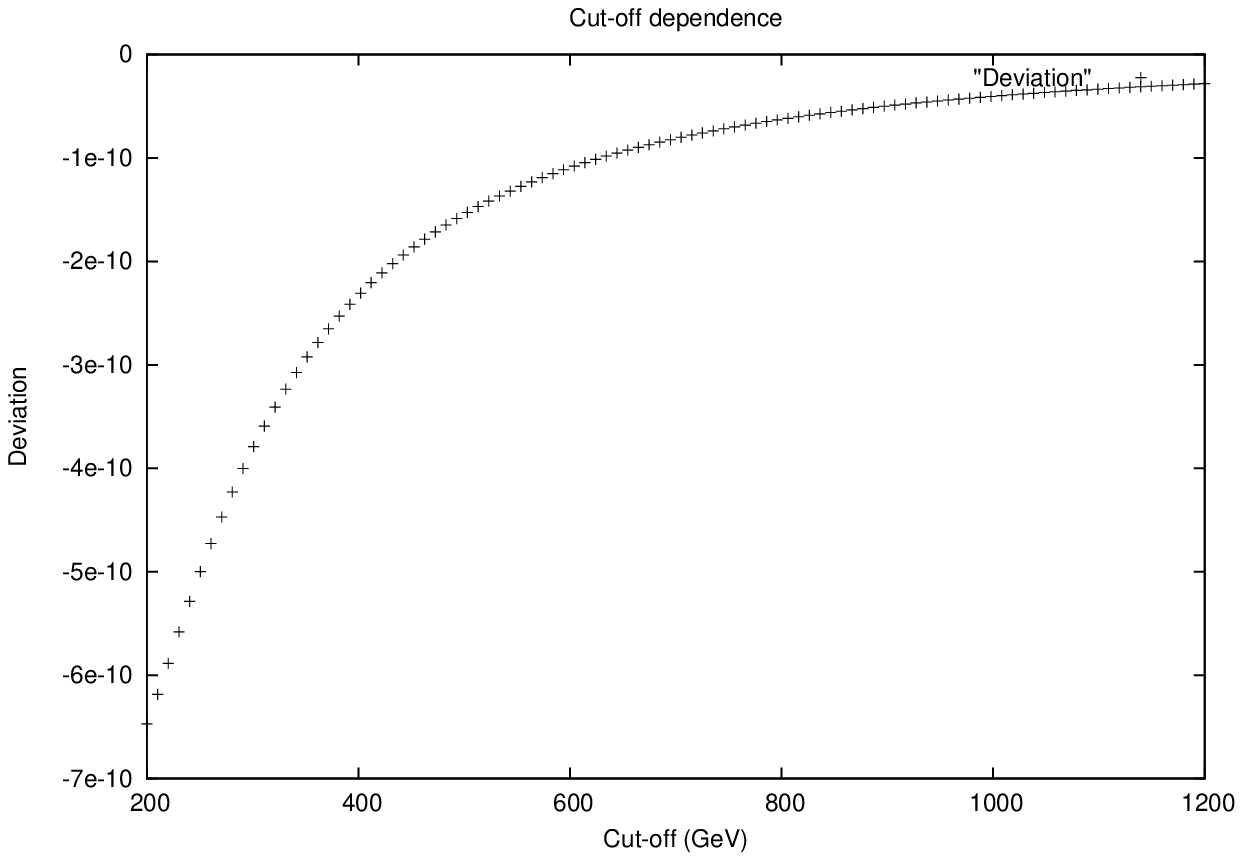, height=90 mm, width=130 mm}

\vspace{10mm}

{\bf Fig. 1: Cut-off ($\Lambda$) dependence of the deviation
$[a_{\mu}^{BY}-a_{\mu}^{SM}](\gamma+W+Z)$.
}
\newline

It is evident from the recent comparison between the experimental and the theoretical SM
prediction in ref. \cite{Blum,Gnendiger}: 

\begin{eqnarray*}
& &a_{\mu}^{1+2\ loops}(Higgs) = +{\cal O}(10^{-11}), \\
& &a_{\mu}^{exp}-a_{\mu}^{SM}=2.7\times 10^{-9},
\end{eqnarray*}

that the loop corrections of the BY theory cannot explain the deviation from the experimental value. 
It seems that more reliable estimates of the hadronic contributions is necessary. They should be 
studied by the nonperturbative methods supplemented by the experiments with hadrons
\cite{Blum}. The new experiments to measure the anomalous magnetic moment ot the muon are planned in
the U.S.A. and Japan with a good potential to further reduce the experimental error.
\newline
\newline

{\bf Appendix}
\newline

The appendix is devoted to the exposure of the explicit formulas for scalar, vector and
tensor functions, as well as for one, two and three point scalar Green functions.

Let us start with definitions and conventions as in \cite{Logan}:

\begin{eqnarray*}
\frac{\imath}{16 \pi^{2}} A(m^{2}) = \int \frac{d^{4}k}{(2 \pi)^{4}}\frac{1}{k^{2}-m^{2}}, 
\end{eqnarray*}
\begin{eqnarray*}
\frac{\imath}{16 \pi^{2}} B_{0;\mu}(q^{2};m_{1}^{2},m_{2}^{2})
 = \int \frac{d^{4}k}{(2 \pi)^{4}}\frac{1;k_{\mu}}{[k^{2}-m_{1}^{2}][(k+q)^{2}-m_{2}^{2}]}, 
\end{eqnarray*}
\begin{eqnarray*}
B_{\mu} = q_{\mu} B_{1},
\end{eqnarray*}
\begin{eqnarray*}
\frac{\imath}{16 \pi^{2}} C_{0;\mu;\mu\nu}(p_{1}^{2},p_{2}^{2},p^{2};m_{1}^{2},m_{2}^{2},m_{3}^{2})
 = \int \frac{d^{4}k}{(2 \pi)^{4}}\frac{1;k_{\mu};k_{\mu}k_{\nu}}
{[k^{2}-m_{1}^{2}][(k+p_{1})^{2}-m_{2}^{2}][(k+p_{1}+p_{2})^{2}-m_{3}^{2}]},
\end{eqnarray*}
\begin{eqnarray*}
C^{\mu} = p_{1}^{\mu}C_{11} + p_{2}^{\mu}C_{12},\ p = -p_{1}-p_{2},
\end{eqnarray*}
\begin{eqnarray*}
C^{\mu\nu} = g^{\mu\nu}C_{24} + p_{1}^{\mu}p_{1}^{\nu}C_{21}
+ p_{2}^{\mu}p_{2}^{\nu}C_{22} + (p_{1}^{\mu}p_{2}^{\nu}+p_{2}^{\mu}p_{1}^{\nu})C_{23}.
\end{eqnarray*}

The standard Green function can be found in ref. \cite{Logan}, while the Green functions in 
the noncontractible space ($\Lambda < \infty$) have the form \cite{Palle2}:

\begin{eqnarray*}
\Re B_{0}^{\Lambda}(p^{2};m_{1},m_{2})=
\frac{1}{2}[\Re \tilde{B}_{0}^{\Lambda}(p^{2};m_{1},m_{2})+
\Re \tilde{B}_{0}^{\Lambda}(p^{2};m_{2},m_{1})],
\end{eqnarray*}
\begin{eqnarray*}
\Re \tilde{B}_{0}^{\Lambda}(p^{2};m_{1},m_{2})=
(\int_{0}^{\Lambda^{2}}d y K(p^{2},y)+\theta (p^{2}-m_{2}^{2}) 
 \int_{-(\sqrt{p^{2}}-m_{2})^{2}}
^{0}d y \Delta K(p^{2},y) )\frac{1}{y+m_{1}^{2}}, \\
K(p^{2},y)=\frac{2 y}{-p^{2}+y+m_{2}^{2}+
\sqrt{(-p^{2}+y+m_{2}^{2})^{2}+4 p^{2} y}},  \hspace*{40 mm}\\
\Delta K(p^{2},y)=\frac{\sqrt{(-p^{2}+y+m_{2}^{2})^{2}+4 p^{2} y}}{p^{2}},
 \hspace*{60 mm}
\end{eqnarray*}

\begin{eqnarray*}
\Re C_{0}^{\Lambda}(p_{1}^{2},p_{2}^{2},p_{3}^{2};
m_{1}^{2},m_{2}^{2},m_{3}^{2}) &=& \frac{1}{3}
[\Re \tilde{C}_{0}^{\Lambda}(p_{1}^{2},p_{2}^{2},p_{3}^{2};
m_{1}^{2},m_{2}^{2},m_{3}^{2}) \\
&+& \Re \tilde{C}_{0}^{\Lambda}(p_{2}^{2},p_{3}^{2},p_{1}^{2};
m_{2}^{2},m_{3}^{2},m_{1}^{2})+
\Re \tilde{C}_{0}^{\Lambda}(p_{3}^{2},p_{1}^{2},p_{2}^{2};
m_{3}^{2},m_{1}^{2},m_{2}^{2}],
\end{eqnarray*}
\begin{eqnarray*}
\Re C_{0}^{\Lambda}(p_{i},m_{j})=\int_{0}^{\Lambda^{2}}dq^{2} \Phi 
(q^{2},p_{i},m_{j})
+\int_{TD}dq^{2} \Xi (q^{2},p_{i},m_{j}), \hspace*{10 mm}\\
\Re C_{0}^{\Lambda}(p_{i},m_{j})=Re C_{0}^{\infty}(p_{i},m_{j})-
\int^{\infty}_{\Lambda^{2}}d q^{2} \Phi (q^{2},p_{i},m_{j}), 
 \hspace*{20 mm} \\
\Phi \equiv function\ derived\ by\ the\ angular\ integration\ 
after\ Wick's\ rotation,\\
C_{0}^{\infty}\equiv standard\ 't\ Hooft-Veltman\ scalar\ function, 
\hspace*{20 mm} \\
TD\equiv timelike\ domain\ of\ integration. \hspace*{40 mm}
\end{eqnarray*}

We list the Green functions (real parts only) necessary for
the one loop contributions with a virtual W boson (photon and Z boson contributions
proceed similarly):

\begin{eqnarray*}
& &C_{0}=C_{0}(m^{2},0,m^{2};0,M_{W}^{2},M_{W}^{2}), \\
& &\lim_{m^{2}\rightarrow 0} C_{11} = \frac{1}{2}[\frac{d B_{0}(m^{2};0,M_{W}^{2})}
{d m^{2}}(m^{2}=0)-C_{0}(m^{2}=0)+M_{W}^{2} \frac{d C_{0}}{d m^{2}}(m^{2}=0)], \\
& &\lim_{m^{2}\rightarrow 0} C_{21} = -\frac{C_{24}(m^{2})}{d m^{2}}(m^{2}=0)
-\frac{1}{2} C_{11}(m^{2}=0) +\frac{M_{W}^{2}}{2}\frac{d C_{11}}{d m^{2}}(m^{2}=0) \\
& &\ \ \ \ \ \ \ +\frac{1}{2}\frac{d B_{1}(m^{2};0,M_{W}^{2})}{d m^{2}}(m^{2}=0), \\
& &\frac{d C_{24}(m^{2})}{d m^{2}}(m^{2}=0)=\frac{1}{4}[C_{11}(m^{2}=0)-M_{W}^{2}
\frac{d C_{11}}{d m^{2}}(m^{2}=0)], \\
& &C_{11}(m^{2})=\frac{1}{2 m^{2}}[B_{0}(m^{2};0,M_{W}^{2})-B_{0}(0;M_{W}^{2},M_{W}^{2})
-(m^{2}-M_{W}^{2})C_{0}], \\
& &B_{1}(m^{2};0,M_{W}^{2})=\frac{1}{2 m^{2}}[-A(M_{W}^{2})+A(0)+(M_{W}^{2}-m^{2})
B_{0}(m^{2};0,M_{W}^{2})], \\
& &B_{0}^{\Lambda}(0;0,M_{W}^{2})=\ln\frac{\Lambda^{2}+M_{W}^{2}}{M_{W}^{2}}, 
\end{eqnarray*}
\begin{eqnarray*}
& &\frac{d B_{0}^{\Lambda}(m^{2};0,M_{W}^{2})}{d m^{2}}(m^{2}=0)
=\frac{1}{2}\frac{1}{M_{W}^{2}}-\frac{1}{4}\frac{M_{W}^{2}}{(\Lambda^{2}+M_{W}^{2})^{2}}, \\
& &\frac{d^{2} B_{0}^{\Lambda}(m^{2};0,M_{W}^{2})}{d (m^{2})^{2}}(m^{2}=0)
=\frac{1}{3}\frac{1}{M_{W}^{4}}+\frac{1}{3}\frac{M_{W}^{2}}{(\Lambda^{2}+M_{W}^{2})^{3}}
-\frac{1}{2}\frac{M_{W}^{4}}{(\Lambda^{2}+M_{W}^{2})^{4}}, \\
& &C_{0}^{\Lambda} (m^{2}) = C_{0}^{\infty} (m^{2}) + \Delta C_{0}^{\Lambda},\ 
C_{0}^{\infty} (m^{2}) = \frac{1}{m^{2}} \ln \frac{M_{W}^{2}-m^{2}}{M_{W}^{2}}, \\
& &\Delta C_{0}^{\Lambda}(m^{2}) = \frac{1}{3 m^{2}}[\int^{1/\Lambda}_{0}d y y^{-1}\frac{1-m^{2}y^{2}}
{(1+M_{W}^{2}y^{2})^{2}}(\sqrt{1+\frac{4m^{2}y^{2}}{(1-m^{2}y^{2})^{2}}}-1) \\
& &\ \ \ +2\int^{1/\Lambda}_{0}d y y^{-1}(1-\frac{y^{2}(M_{W}^{2}-m^{2})+1}
{\sqrt{(1+y^{2}(M_{W}^{2}-m^{2}))^{2}+4m^{2}y^{2}}})],
\end{eqnarray*}
\begin{eqnarray*}
& &C_{0}^{\Lambda}(m^{2}=0)=\frac{1}{M_{W}^{2}} (-1+x(1+x)^{-1}),\ x=\frac{M_{W}^{2}}{\Lambda^{2}}, \\
& &\frac{d C_{0}^{\Lambda}}{d m^{2}}(m^{2}=0)=\frac{1}{M_{W}^{4}}[-\frac{1}{2}
+\frac{2}{3}(-(1+x)^{-2}+(1+x)^{-3})], \\
& & \frac{d^{2} C_{0}^{\Lambda}}{d (m^{2})^{2}}(m^{2}=0)=\frac{1}{M_{W}^{6}}[-\frac{2}{3}
+\frac{4}{3}(-2(1+x)^{-5}+8(1+x)^{-4}-\frac{37}{3}(1+x)^{-3} \\
& &\ \ \ +9(1+x)^{-2}-3(1+x)^{-1}+\frac{1}{3})].
\end{eqnarray*}

\end{document}